\documentclass[12pt]{article}
\usepackage{graphicx}
\usepackage{amsthm,amsmath,amsfonts}
\usepackage{csquotes}
\usepackage{lineno}
\usepackage{multirow}
\usepackage{graphicx,subcaption}
\usepackage{amssymb}
\usepackage{enumitem}
\setlist{leftmargin=12.5mm}
\usepackage{makecell}
\usepackage{caption}
\usepackage{floatrow}
\newcommand*{\Scale}[2][4]{\scalebox{#1}{$#2$}}

\usepackage[OT1]{fontenc}
\usepackage{amsthm,amsmath,amsfonts}
\usepackage[numbers]{natbib}
\setcitestyle{numbers,open={[},close={]}}
\usepackage[colorlinks,citecolor=blue,urlcolor=blue]{hyperref}
\usepackage[a4paper, total={6.5in, 8.5in}]{geometry}
\usepackage{latexsym,graphicx,subcaption,verbatim,nameref}
\usepackage{diagbox}
\usepackage[affil-it]{authblk}
\usepackage[all,cmtip]{xy}
\usepackage[toc,title,titletoc]{appendix}

\usepackage{threeparttable,booktabs}
\usepackage{makecell}
\usepackage{csquotes}
\usepackage{chngcntr}
\usepackage{subcaption}
\usepackage{graphicx}
\usepackage{esint}
\usepackage{relsize}
\usepackage{color}
\usepackage{multirow}
\usepackage{algorithm}
\usepackage{algpseudocode}
\usepackage{setspace}
\onehalfspace
\numberwithin{equation}{section}
\theoremstyle{plain}

\theoremstyle{definition}

\usepackage{mathtools}

\title{Complexity of Government response to Covid-19 pandemic: A perspective of coupled dynamics on information heterogeneity and epidemic outbreak
}

\date{}

\author[1 2]{Xiaoqi Zhang%
\thanks{Email: \texttt{xiaoqizh@buffalo.edu}}}
\author[2]{Jie Fu%
\thanks{Email: \texttt{jz0429@163.com}}}
\author[2]{Sheng Hua%
\thanks{Email: \texttt{huasheng@seu.edu.cn}}}
\author[3]{Han Liang%
\thanks{Email: \texttt{hanliang1@whu.edu.cn}}}
\author[4]{Zi-Ke Zhang %
\thanks{Email: \texttt{zhangzike@gmail.com}; Corresponding Author.}}

\affil[1]{Institute of Economics, Chinese Academy of Social Sciences, Beijing, China}
\affil[2]{National School of Development, Southeast University, Nanjing, China}
\affil[3]{Dong Fureng Institute of Economic and Social Development, Wuhan University, Wuhan, China}
\affil[4]{College of Media and International Culture, Zhejiang University, Hangzhou, China}

\begin{document}

\maketitle

\begin{abstract}

This study aims at modeling the universal failure in preventing the outbreak of COVID-19 via real-world data from the perspective of complexity and network science. Through formalizing information heterogeneity and government intervention in the coupled dynamics of epidemic and infodemic spreading; first, we find that information heterogeneity and its induced variation in human responses significantly increase the complexity of the government intervention decision. The complexity results in a dilemma between the socially optimal intervention that is risky for the government and the privately optimal intervention that is safer for the government but harmful to the social welfare. Second, via counterfactual analysis against the COVID-19 crisis in Wuhan, 2020, we find that the intervention dilemma becomes even worse if the initial decision time and the decision horizon vary. In the short horizon, both socially and privately optimal interventions agree with each other and require blocking the spread of all COVID-19-related information, leading to a negligible infection ratio 30 days after the initial reporting time. However, if the time horizon is prolonged to 180 days, only the privately optimal intervention requires information blocking, which would induce a catastrophically higher infection ratio than that in the counter-factual world where the socially optimal intervention encourages early-stage information spread. These findings contribute to the literature by revealing the complexity incurred by the coupled infodemic-epidemic dynamics and information heterogeneity to the governmental intervention decision, which also sheds insight into the design of an effective early warning system against the epidemic crisis in the future.
\end{abstract}

\section{Introduction}
\label{intro}
\begin{figure}
    \centering
\includegraphics[width=9cm,height=10cm]{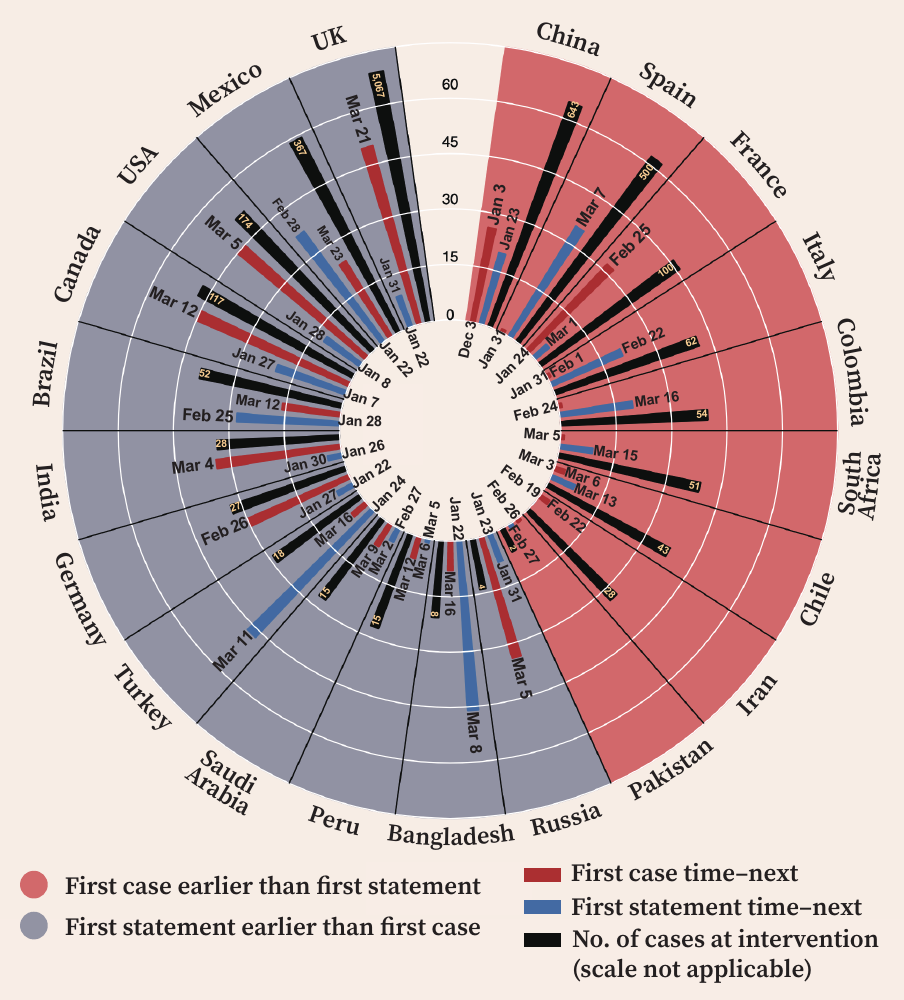}
\caption{Timeline of COVID-19 outbreak and government responses in 21 countries}
\floatfoot{\footnotesize Timelines of COVID-19 outbreak in the top 20 countries ranked according to the number of their cumulative infectious cases of COVID-19 by Jun. 30, 2020, and China, the country with the first confirmed infectious case of COVID-19. The event time data comes from Wikipedia \cite{wiki}, and the infection number comes from the COVID-19 dashboard at the John Hopkins University \cite{johnhopkins}. Using the blue and red color bar, we mark three milestones on the timeline of each country that are the time when the first confirmed case was reported, the time when the central government first published an official statement regarding COVID-19, and the time when the country first seriously executed non-pharmaceutical intervention. The left half of the annular (with blue background) corresponds to those countries that have their first official statement made ahead of the first domestic infectious case, and the right half of the annular (with red background) corresponds to the countries that have their first domestic case reported ahead of the official statement. The time difference (days) from the first domestic infectious case to the next event (the first official statement or the first execution of NPIs) is measured by the length of the red bar, while the time difference from the first official statement to the next event (the first domestic cases or the first NPIs) is measured by the length of the blue bar. The two-time differences measure the speed of government response toward the outbreak of COVID-19. For each country, the exact occurrence date of every event is marked at the ends of the blue and red bars with the date marked toward the center of the annular representing date of the earlier event. We also sketch the number of cumulative infectious cases at the time of the first NPIs with the black bar and mark the exact number on the top of the black bar. The comparison of the relative length of three color bars across countries provides a measure for the relative speed of COVID-19 spreading versus the government response.}\label{timeline}
\end{figure}

Since the outbreak of COVID-19, a variety of major national governments have failed to manipulate preventive strategies against the pandemic one after another. As shown in the Fig. \ref{timeline}, no matter whether the countries made a fast official warning to the public, whether their first domestic case of COVID-19 was reported before or after the government's warning statement, whether or not the government had a strict restriction on its domestic information spreading media, almost all countries suffered from the failure of controlling the pandemics of COVID-19 \cite{zhan2022explore}. Fig. \ref{timeline} suggests a counter-example to the theory that free information spreading and fast warning are helpful in containing the outbreak of an epidemic crisis. We believe that the worldwide failure in front of the COVID-19 crisis reflects some systematic incapability of tackling the unknown infectious diseases by modern government; one source of the incapability is the policy dilemma: remedy or overkill.

On the one hand, the government may choose an inactive move (remedy), such as blocking information spreading in the early stage of an epidemic crisis so as to keep the public away from panic, by which the infection ratio may not grow too fast in the early stage while the government suffers the least public pressure. On the other hand, the government can choose a socially optimal strategy (overkill), by which the government encourages information spreading and even discloses as much classified information to the public as possible. By doing so, a fast accumulation of infectious cases is witnessed by the public, which may effectively send warning signals but also increase public pressure, therefore, is not optimal from the government's angle. The conflict between the government's private optimal and social optimal makes it almost unlikely for the government to make the right decision for the public, which triggers the outbreak of the public health crisis. Therefore, we need to understand how this conflict arises during an epidemic crisis, and whether and how it is avoidable. To this end, information is one key factor as it is believed to be critical for the early-stage control of any infectious disease.

It is demonstrated that the key factor in controlling the spread of epidemics successfully is to understand the complex two-way interaction between disease dynamics and human social behavior\cite{wang2015coupled}. First, the epidemic spreads can stimulate information spreading about the disease, leading to people’s awareness of the crisis\cite{lynch1991thought,tai2007media,pastor2015epidemic}. Some people will take preventive measures to protect themselves from the disease\cite{ferguson2007capturing,ruan2012epidemic}. From this perspective, the epidemic dynamics experience an endogenous negative feedback\cite{granell2013dynamical,funk2009spread,funk2010modelling,sahneh2012existence,lima2015disease}, by which the disease might be self-contained. Second, information propagation may not always induce preventive behavior; it can oppositely stimulate more risky behavior and accelerate the spread of disease\cite{kasperson1988the}. The positive feedback exists because information does not only come from people's own observations and experiences (source information), but also mingles unreliable second-hand information from social media\cite{starbird2010pass,murakami2012tweeting,starbird2014rumors}, where people can get information and interactive views\cite{burnap2014tweeting,vieweg2010microblogging}. On social media, there is a spontaneous trend of turning information into rumors with strong emotional tendency\cite{davis2020phase,jones2017distress}, which increases anxiety feelings and magnifies irrational behavior\cite{taha2014h1n1,bordia1998rumor,ng2018to,slovic1987perception}. As an example, it is observed that 41.3\% of diagnosed patients of COVID-19 have contracted the disease during hospital visits in Wuhan \cite{wangdawei2020}.

As the main body of social governance, the government is also the most critical information node in the entire social network. Although the government is often asked to disclose the \enquote{truth} to the public as soon as possible, it shares the same kind of uncertainty as the public when facing such an unprecedented virus as COVID-19, there is no \enquote{truth} to disclose at least at the early stage. Restricted by this real constraint, the effectiveness of any government responses may be questionable. Therefore, this study aims to theorize the role of government intervention within a coupled infodemic-epidemic dynamic model where both positive and negative feedback are allowed. We further conducted the counterfactual analysis on both the synthetic settings and the parameters calibrated from real-world pandemic data in order to reveal the heterogeneous impact of information on epidemic spreading and its incurred decision complexity faced by government. We believe they are critical to the outbreak of the COVID-19 pandemic, hence should be paid more attention in the future design of warning systems for a public health crisis.

The remaining sections are organized as the following. Section 2 introduces the coupled dynamic model of information and epidemic spreading and formalizes the government intervention decision problem through the model. Section 3 analyzes the mechanism that leads to the government response dilemma by numerical experiments on both the synthetic model setting and the calibrated setting derived from the early-stage epidemic data of Wuhan, China. We also briefly discuss the potential solution to the government response dilemma from the perspective of an information network. Section 4 concludes the findings and discusses the direction of future research.

\section{Literature Review}
\label{sec:2}
\subsection{Infodemic and Epidemic Spreading}
\label{sec:2.1}
The current study is closely related to the literature on the co-evolution dynamics of epidemics and infodemics \cite{li2023coevolution,hong2022co,zhang2021multiplex}. It has been widely acknowledged that the complex two-way interaction between disease dynamics and human social behavior is critical to the outbreak of an epidemic crisis and its prevention. The information on the epidemic may induce fear effect \cite{debnath2023memory} and intrigue self-protection behavior, which helps contain the outbreak of epidemic crisis  \cite{debnath2023memory,wu2022effect}. Despite the positive effect, studies \cite{kreps2020model,cinelli2020covid,gu2022influence,gisondi2022deadly} found that information may not always be kept real, mis-information is inevitable and will reduce the scientific trust of the public, which incurs negative impact on the containment of epidemic crisis. More recent studies jump out of the dichotomy between information and mis-information, the heterogeneity in information literacy \cite{wu2022effect}, sources \cite{hong2022co} channels by which information is propagated \cite{zhang2021multiplex}, and their impact on the epidemic spreading are thoroughly investigated.

Despite the persistent discussions on the co-evolution of infodemic and epidemic, the focus of the literature was inclined to theoretically model the co-evolution process itself and/or empirically estimate the extent of infodemic-epidemic interaction. It was rarely studied whether and how the co-evolution process can be intervened and/or utilized for governance purposes. Particularly, whether or not the control of information propagation by the government can be supported by the coupled infodemic-epidemic dynamics \cite{zhan2018coupling} and facilitate the containment of the outbreak of the epidemic crisis is still an open question, to which we attempt to give a formal discussion.

\subsection{Prevention and Control}
Our study is also connected to the literature on the optimal control and prevention strategies against epidemic crisis \cite{zhang2020evaluating,ghosh2021qualitative,mondal2022mathematical,ghosh2022mathematical,majumdar2022controlling}. Existing studies have extensively discussed a long list of the preventive measures, including but not limited to the social distancing \cite{ghosh2022mathematical}, traffic jam and lockdown \cite{zhang2020evaluating,zhou2023psychological}, vaccination \cite{ghosh2021qualitative}, mask-use \cite{bartsch2022maintaining}, and the effectiveness and potential trade-offs incurred by them. It is worthwhile noting that these discussions focus almost exclusively on the late-stage of the epidemic crisis. Unless the outbreak of infectious diseases has persisted for long or the spreading cannot be contained, it is not really needed to adopt any social distancing measure \cite{sun2022effectiveness}. On the the other hand, during the early-stage of an epidemic outbreak, the lack of knowledge and awareness on infectious diseases, such as COVID-19 \cite{lu2021re}, makes it also reluctant to adopt any prevention measure in the aforementioned list. From this perspective, information might be the only way in the early-stage epidemic crisis that the government can intervene to affect and guide the public to protect themselves \cite{lu2021re}. However, to the information intervention and its effectiveness and trade-offs in the early-stage coupled dynamics of infodemic and epidemic crisis, the existing studies paid very little attention, to which a formal investigation will be carried out in this study.

From the brief review, we identify our main contribution to the literature as bridging the two branches of research on the coupled infodemic-epidemic dynamics and the optimal intervention, which is critical to the early-stage warning and prevention against an epidemic crisis.
\color{black}
\section{Methods}
\subsection{SI-NLH model: set-up and properties}

To capture the heterogeneous response to different types of information and the induced impact on disease dynamics, in this section, we propose an SI-NLH dynamic model where the SI component
is the classical Susceptible-Infection model capturing the disease dynamics. The NLH component reflects the dynamic transition of individuals among three information statuses: {\bf N}o information, with {\bf L}ow-quality information and {\bf H}igh-quality information.

Compared with the SIR/SIS model, the SI model excludes the recovery dynamics and is, therefore, more appropriate to model the early-stage disease dynamics in the real world, such as the pandemic of SARS and COVID-19. As in the early stage of epidemic spreading, the number of recovered cases is completely negligible in relation to the newly infected cases\cite{zhou2006behaviors}. In fact, a rigorous mathematical analysis\cite{bertozzi2020challenges} shows that when the infection/recovery ratio, i.e., the ratio of infected/recovered cases versus the population size, is very small, the infection dynamics derived from a SI model and a SIR model are equivalent. At the early stage of the epidemic crisis, the small infection/recovery ratio always holds (for instance, in the COVID-19 pandemic in China, even in the epicenter, Wuhan, the greatest infection/recovery ratio has never exceeded 0.8\%); therefore, it is safe to consider the simpler SI model for the epidemic spreading. (In the supplementary material to this study, we conduct a simulation study similar to the one discussed in the following sections, but based on a SIR-NLH model where the SI component is replaced with the SIR model, the result is similar, the details can be found in Supplementary Fig. 1.)

The differentiation of information quality in the NLH component is critical, as the quality often determines the response behavior of individuals toward infectious diseases. High-quality information consists of that which is officially issued by truth holders, such as the government or specialists in infectious diseases. The contents of this information are correct and the description is complete, therefore they can intrigue self-protection behavior and lower the infection risk. In contrast, low-quality information is rumors and/or that whose contents are not completely wrong but the description is incomplete. This information can easily trigger incorrect interpretations and be converted to rumors in later-stage propagation. Consequently, low-quality information is likely to intrigue panic and irrational behaviors that rise up the infection risk\cite{GallottiValle-6399}.
However, the existing literature on coupled infodemic-epidemic dynamics focuses on the positive side of the information \cite{wang2015coupled,wang2016suppressing,kim2019incorporating}, the low-quality information and the induced higher infection risk are often missed. Although some recent studies start to pay attention to the negative side of information\cite{zhang2021multiplex,lu2021re}, they mainly focus on empirical and/or simulation studies on the epidemic consequences of rumor spreading rather than theoretically modeling the underlying mechanism. In the SI-NLH model, we integrate both the positive and negative sides of information and their interaction mechanism, which provides a more unified view of the coupled infodemic-epidemic dynamics.

\begin{figure}
    \centering
\includegraphics[width=8.5cm,height=3.5cm]{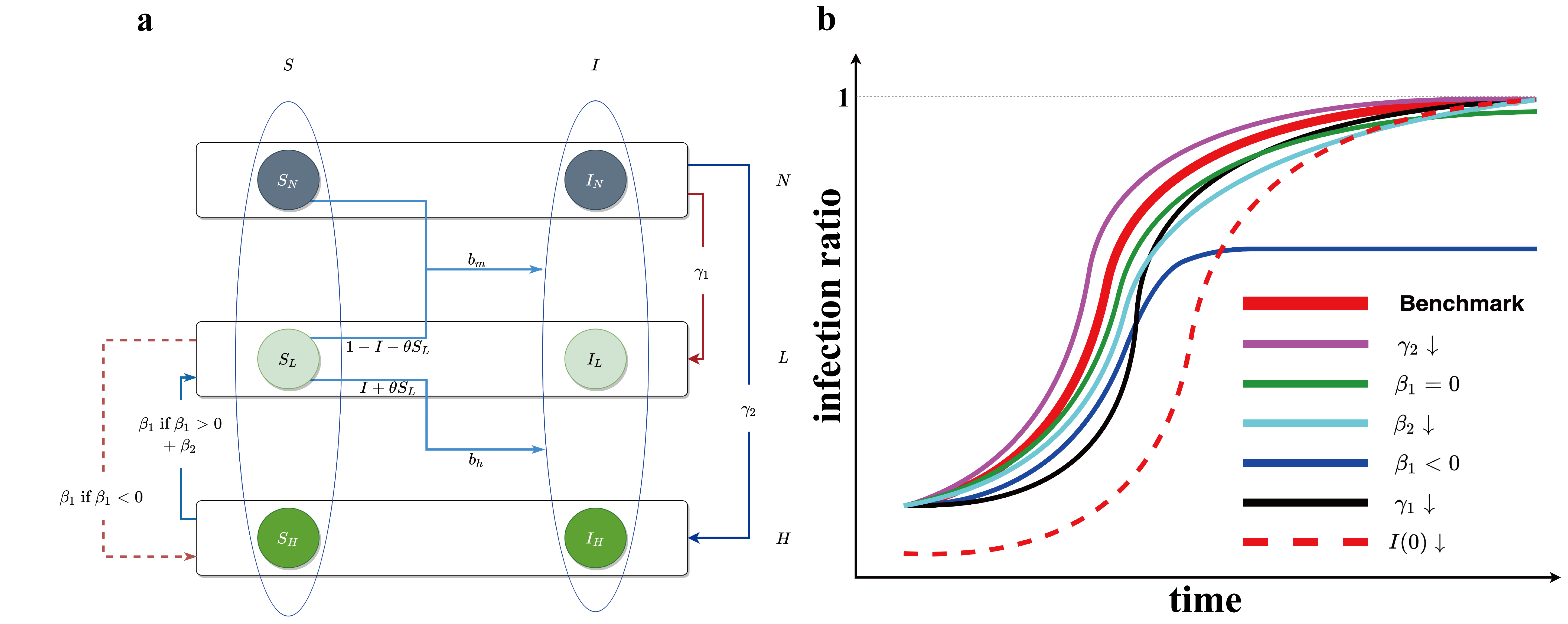}
    \caption{Coupled SI-NLH model and trajectories of infection ratios generated under different information parameters}
    \floatfoot{\footnotesize {\bf (a)} Illustration of the six compartments and their interconnection within the SI-NLH model. {\bf (b)} Evolution of infection ratio for a variety of information parameters and initial conditions. Given the benchmark trajectory of the infection ratio growth (bold red), trajectories are plotted against a variety of modifications on the model set-up, including i) a decrease in the initial infection ratio $I(0)$ (dashed red); ii) decrease in the information transition rate $\gamma_1$ (black), $\gamma_2$ (purple) and the decay rate $\beta_2$ (light blue); iii) high-quality information dominates the information dynamics, $\beta_1=0$ (green) and $\beta_1<0$ (blue).}\label{toy}
\end{figure}

According to the joint of the information and disease status, the population in the SI-NLH model is divided into six groups, namely the susceptible without information ($S_N$), the susceptible with low-quality information ($S_L$) and the susceptible with high-quality information ($S_H$), the infected without information ($I_N$), the infected with low-quality information ($I_L$) and the infected with high-quality information ($I_H$). The fraction of the six groups and their dynamic transitions are formally described via the following ordinary differential equation system and illustrated in Fig. \ref{toy}a,
\begin{equation}\label{toy math}
\begin{cases}
S_N'=&-\gamma_1S_N L (1-I)-\gamma_2S_N H(1-I)-b_m S_N I\\
S_L'=&\gamma_1 S_N L\cdot(1-I) +S_H(1-I)(\beta_1 L+\beta_2 )\\&-b_h S_LI(I+\theta S_L)-b_mS_LI(1-I-\theta S_L)\\
S_H'=&-S_H(1-I)(\beta_1 L+\beta_2 )+\gamma_2 S_NH(1-I)   ,\\
I_N'=&-\gamma_1I_N L -\gamma_2I_N H+b_m S_N I\\
I_L'=&\gamma_1 I_N L +I_H(\beta_1 L+\beta_2 )+b_h S_L I(I+\theta S_L)\\&+b_m S_L I (1-I-\theta S_L)\\
I_H'=&-I_H (\beta_1 L+\beta_2 )+\gamma_2 I_N H
\end{cases}
\end{equation}
where the fractions $L$, $H$, $I$ and $S$ satisfy the relations that $L=I_L+S_L$, $H=I_H+S_H$, $I=I_N+I_L+I_H$, $S=S_N+S_L+S_H$ and $I+S_N+S_L+S_H\equiv 1$. Based on Eq. \eqref{toy math} and Fig. \ref{toy}a, there are three classes of parameters governing the SI-NLH dynamics, which are i) information parameters: $\gamma=(\gamma_1,\gamma_2)$, characterizing the rate of transition from No information to Low- and High-information, and $\beta=(\beta_1,\beta_2)$, the rate of transition between L- and High-information; ii) the transmission rates to disease $b=(b_m,b_h)$ for susceptible people under unawareness ($b_m$) and panic ($b_h$), respectively; and iii) the panic parameter $\theta$ quantifying the susceptibility to panic. For the convenience of reading, a full list of the notations and meanings for our model parameters and endogenous variables are presented in Appendix Table \ref{table: table1}.

For the three classes of parameters, we made the following assumptions in order to reflect the real-world interaction mechanism between infodemic and epidemic spreading. First, for information parameters, we suppose that information spreading is irreversible, which means the rate $\gamma_1$/$\gamma_2$ of $N$-type individuals transmitting to $L$-/$H$-type is nonnegative. For the transmission between $L$-type and $H$-type, we let the sign of $\beta_1$ characterize which direction of transmission is dominant, a positive $\beta_1$ implies the transmission from $H$ to $L$ is prevalent, the negative $\beta_1$ implies the contrary. To capture the natural trend of the information decay during propagation, we set a nonnegative rate $\beta_2$ to capture the transmission from $H$ to $L$ due to quality decay. Second, for disease dynamics, we assume that individuals with different types of information would react differently to the disease, which leads to different transmission rates. $S_H$ owns high-quality information and tends to self-protect which leads to a low transmission rate, which is set to $0$ for the sake of simplicity (this assumption does not significantly affect the result of this paper, we relax it in the supplementary material and report the simulation results in supplementary Fig. 2 in which $S_H$ is allowed be infected in a positive rate $b_l$ that is lower than the rate of the $S$ individuals with the other types of information). Because of no information, $S_N$ is unaware of the disease and likely to maintain their regular behavioral patterns, which yields a middle-level transmission rate $b_m>0$. $S_L$ behaves more complicatedly. On one side, they intend to ignore their information due to its low quality. In this case, they behave as if being unaware of the disease, yielding the transmission rate $b_m$. On the other side, the low quality might stimulate panic or other kinds of irrational behaviors for $S_L$, such as gathering in hospitals and supermarkets, which leads to a much higher infection risk $b_h \gg b_m$. Finally, panic during an epidemic crisis always results from the increasing infectious cases and the self-exciting property that the more people already in a panic mood, the more likely they are to entail panic among the remaining people. Therefore, we assume $S_L$ has the probability $I+\theta S_L$ of turning to panic, where $\theta$ is the panic parameter measuring the self-exciting degree of panic.

Based on the aforementioned model set-up, the disease propagation following a SI model implies that as $t\rightarrow\infty$ the infection ratio $I(t)$ would always converge, while the limit ratio depends on the choice of information parameters $\beta$. As shown in Fig. \ref{toy}b, the ultimate infection ratio would almost always converge to 1, i.e. the entire population will be infected, provided that $\beta_1\geq -\beta_2$, no matter where the dynamic system starts. On the contrary, when $\beta_1<-\beta_2$, the limit infection ratio is possible to be strictly less than $1$, the limit depends on both the initial condition and the parameter $\beta_1$ and $\beta_2$. Especially, when the configuration is fixed, the infection ratio increases with both $\beta_2$ and $\beta_1$. The remaining information parameter $\gamma$, the transmission rate, and the panic parameter have no impact on the ultimate infection ratio, but would significantly impact the convergence speed as shown in Fig. \ref{toy}b.
In particular, it can be proved that the temporal infection ratio $I(t)$ is monotonically increasing with the parameter $\gamma_1$, $\beta_1$, $\beta_2$ and decreasing with parameter $\gamma_2$ when all others keep constant (Fig. \ref{toy}b shows a few examples of the general relation).

Fig. \ref{toy}b shows a different perspective for thinking of information governance during the early stage of epidemic spreading. Once the heterogeneity of information is introduced, it is no longer correct as claimed in the literature \cite{granell2013dynamical,funk2009spread,funk2010modelling,sahneh2012existence,lima2015disease} that more information can help better contain disease outbreaks. The low-quality information cannot help mitigate the infection risk, but even increase the risk exposure of susceptible individuals. The heterogeneity of information is neither static nor external to the disease dynamics, but co-evolves with it. This fact makes the underlying structure of the information market extraordinarily important because it shapes the propagation process and determines which type of information will be prevalent and which type will be thrown away. The information filtered by propagation will feed back again to affect disease dynamics. In the situation that $\beta_1>0$ (the solid red line in Fig. \ref{toy}b) and/or $\beta_1=0$ meanwhile $\beta_2>0$ (solid green line in Fig. \ref{toy}b), an information market admitting free propagation of all information types would intrinsically squeeze out the high-quality information and gradually make all people exposed to the rumors and diseases. In this situation, government intervention in the information market will become indispensable.

\subsection{Government intervention in SI-NLP model: classification and trade off}
During the disease pandemic in the real world, such as the pandemic of COVID-19, it is often the case that information about the disease passes through a hierarchical network in the direction from the bottom to the top. As to be discussed in later sections, such a network structure and information flow direction makes information quality decay inevitable which further stimulates the outbreak of disease. Therefore, external intervention from an independent third-party is indispensable. Government can intervene in the infodemic-epidemic co-evolution process to achieve a low ultimate infection ratio. However, it is not so easy for the government to figure out the correct time and the correct intervention methods. To better interpret the trade-off and dilemma faced by the government, we theorize the role of government intervention in the SI-NLH model.

To be more concentrated, we focus only on the intervention measures on information dynamics, which means the government can only have an impact on the values of the information parameters $\gamma$ and $\beta$. This assumption is reasonable as the transmission rates and the panic parameter are determined essentially by the virus and the cultural background of a society that is out of control by the government. Without loss of generality, we concentrate on two classes of intervention strategies: (i) information injection strategy\cite{terpstra2009does,garfin2020novel}; and (ii) the blocking strategy\cite{huo2011interplay,mukkamala2018role}.

\begin{figure}
    \centering
\includegraphics[width=0.9\textwidth]{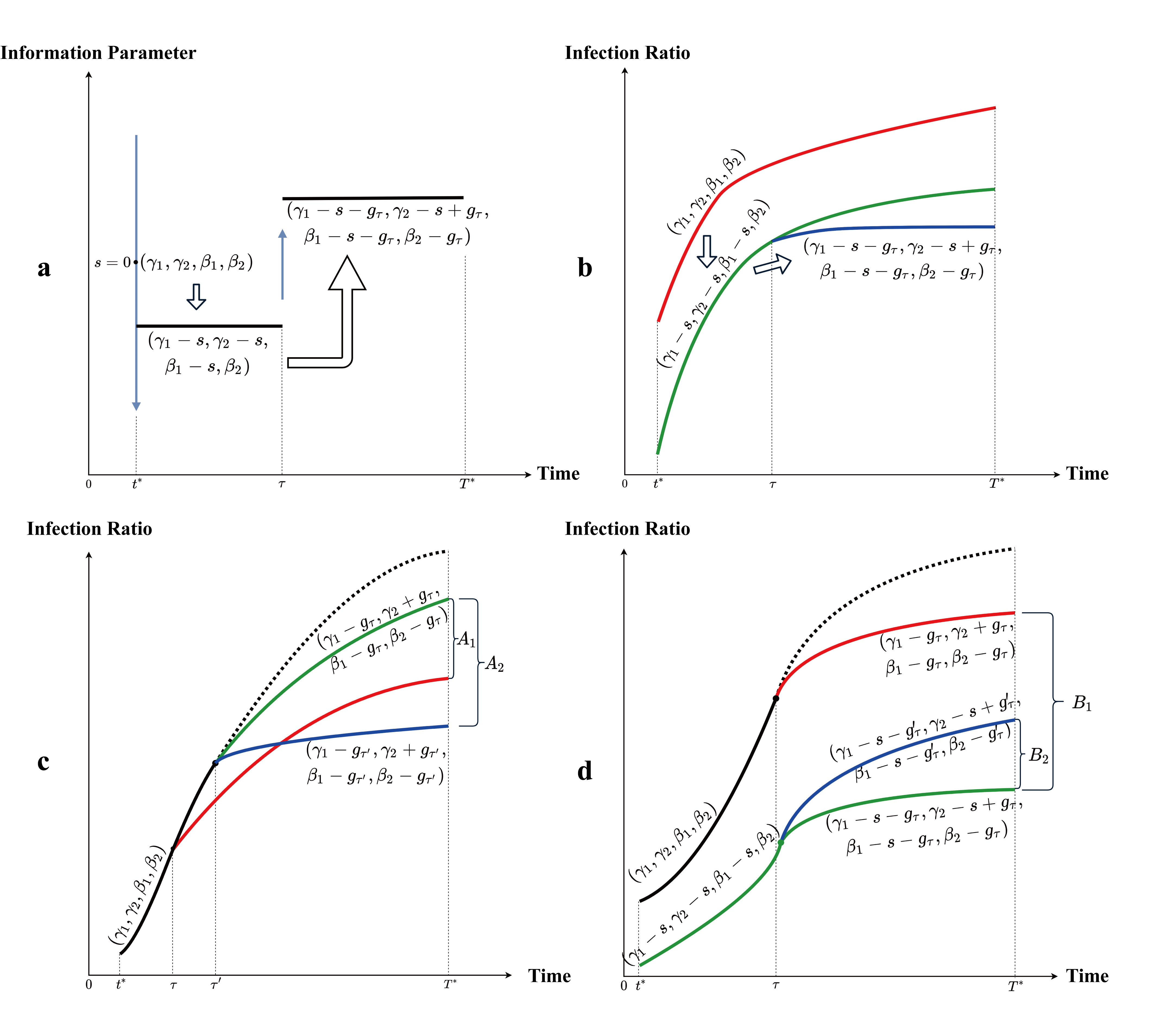}
    \caption{Effect of blocking and information injection.}\floatfoot{\footnotesize  {\bf (a,b)} Illustration of the mechanism that blocking and information injection influence the coupled infodemic-epidemic dynamics. Plot {\bf a} shows their direct impact on information parameters. Plot {\bf b} illustrates their ultimate impact on the trajectory of infection ratio.  {\bf (c,d)} Trade-offs incurred by delayed information injection and blocking. Plot {\bf c} reveals the decomposition of positive ($A_2$) and negative ($A_1$) impact of delayed information injection on suppressing infection ratio, plot {\bf d} decomposes the positive ($B_1$) and negative ($B_2$) impact induced by increasing the degree of blocking.
    }\label{intervention-tradeoff}
\end{figure}

Information injection and blocking impact the coupled dynamics in different ways, which are summarized in Fig.\ref{intervention-tradeoff}a. By injecting more high-quality information into the public, the government can accelerate the conversion of $S_N$ and $S_L$ to $S_H$ and avoid the decay of high-quality information $H$ to $L$, which lowers the population-level transmission rate into infection, resulting in that the infection ratio trajectory decreases from the green-line level to the blue-line level in Fig. \ref{intervention-tradeoff}b. However, before the government can provide high-quality information for the public, it takes time to collect and preprocess tons of noisy information. The easier the government can collect highly accurate information about the disease, the more effective the injection can be, which implies the execution and effectiveness of information injection rely on time and accumulation of information and particularly the high-quality information in the system. In contrast, by information blocking, government controls the speed of information spreading in the system, while does not have to provide any processed information for the system. Therefore, blocking can slow down the rumor spreading which helps lower the population-level transmission rate to infection, resulting in the infection ratio trajectory decreasing from the red-line level to the green-line level in Fig. \ref{intervention-tradeoff}b. However, on the other hand, without extra information being provided, blocking cannot stop the decay of high-quality information, while its effectiveness does not depend on the execution time and the accumulation of information in the system.

To formalize the difference between information injection and blocking in the SI-NLH model, we assume that the epidemic crisis breaks out at time $t^\ast$ since then the government decides whether or not and to what degree to inject and block information. The government made the intervention decision in an attempt to minimize the cumulative impact of the epidemic crisis by a decision time horizon $T^\ast$ with $T^\ast>t^\ast$. To reflect the time-dependence of information injection, we assume it takes effect since a given injection time $\tau\in [t^\ast,T^\ast)$ and induces the increase in rate $\gamma_1$ and decrease in rate $\gamma_2$, $\beta_1$ and $\beta_2$ during the time interval $(\tau,T^\ast]$ by the degree $g_\tau=g(H_\tau, L_\tau+H_\tau)$ as shown in Fig. \ref{intervention-tradeoff}a. To reflect the information-dependence, we let $g_\tau=g(H_\tau, L_\tau+H_\tau)$ monotonically increase in both the ratio of high-quality information $H_\tau$ and the ratio of all informed individuals $L_\tau+H_\tau$. Since the impact of information injection depends on the injection time $\tau$ via the degree function $g_\tau$, the government needs to select the injection time so as to optimize the intervention effect.

For information blocking, we suppose that it induces the decrease in the absolute value of parameter $\gamma_1$, $\gamma_2$ and $\beta_1$ to the degree $s$ since the very beginning $t^\ast$ of the epidemic crisis, but cannot affect the decay rate $\beta_2$, as shown in Fig. \ref{intervention-tradeoff}a. In addition, we suppose that blocking can take effect only if it can be maintained for a while, frequent changes in the blocking degree $s$ would interrupt the normal order in the information market and are forbidden by assumption. Without loss of generality, we assume the degree $s$ of the information block is held constant for the entire period from $t^\ast$ to $T^\ast$. To generalize the scope of analysis, we allow the blocking degree $s$ to take zero and even negative values, which corresponds to no blocking strategy the government does not block any information at all and maintains the current information spreading speed, and the \enquote{negative} blocking strategies by which government attempts to accelerate information spreading among the public.

Both information injection and blocking can generate positive and negative impacts on the control of the final infection ratio, $I(T^\ast)$, at the decision horizon $T^\ast$. Fig. \ref{intervention-tradeoff}c and Fig. \ref{intervention-tradeoff}d sketch those conflicted impacts in detail. In Fig. \ref{intervention-tradeoff}d, the effect of information injection depends on the injection time, a delayed injection from $\tau$ to $\tau'>\tau$ firstly induces a positive structural effect associating with a greater $g_\tau$ which leads to a faster transition from $S_L/S_N$ to $S_H$ and a lower ultimate infection ratio, which is represented as the gap term $A_1$. On the other hand, as shown in Fig. \ref{intervention-tradeoff}c, the later injection makes the infection ratio grow in its original speed during the period $[\tau,\tau')$ that is faster than the speed induced by injection at $\tau$. Although the faster growth of infectious cases is temporal, it increases the cumulative infection ratio by time $\tau'$, which further increases the transmissibility of the disease after $\tau'$ due to a greater initial infectious population at $\tau'$. This negative temporal effect is represented as the term $A_2$ as shown in Fig. \ref{intervention-tradeoff}c.

In Fig. \ref{intervention-tradeoff}d, the impact of blocking on $I(T^\ast)$ is also decomposed into two conflicted parts. First, a positive structural effect that is purely attributable to the change of information parameters, represented by the gap $B_1$, which
slows down the transition to $S_L$ from $S_H$ and $S_N$, therefore suppresses the panic-induced infection risk and lowers the overall infection ratio in the end. Second, the negative injection effect, $B_2$, by which the decrease in information spreading speed would reduce the portion of informed people, represented as $H_\tau$ and $L_\tau+H_\tau$, for every time $\tau$ after the initial decision time $t^\ast$ for blocking that further reduces the effectiveness of information injection, $g_\tau$, and facilitate the growth of ultimate infection ratio.

Note that the four marginal effects, $A_1$, $A_2$ and $B_1$, $B_2$, are not separable in practice, combining them together would disturb the coupled infodemic-epidemic dynamics in a complicated way that induces a joint trade-off effect for the government when facing strategy selection.

\subsection{Set-ups for synthetic analysis and model calibration}
We will numerically analyze the effect of both intervention strategies on the final infection ratio for a wide range of synthetic settings. For fixed initial decision time $t^\ast$ and decision horizon $T^\ast$, we will calculate the variation trend of the intervention effect (measured by $I(T^\ast)$, the infection ratio at the end of the decision horizon) along with the joint adjustment of injection time $\tau$ and blocking degree $s$. We let $\tau$ range within $(t^\ast,T^\ast)$ and let $s$ vary within $[0,2]$ such that the three information parameters are determined via $\gamma_1=\gamma_2=\beta_1=s$. We also test the sensitivity of our analytic results against the change of both $t^\ast$ and $T^\ast$, for which
the {\bf benchmark} setting is represented by the initial distribution of population within the six groups $(S_N,S_L,S_H,\\I_N,I_L,I_H)=(0.99,0,0.009,0.001,0,0)$ and the decision horizon $T^\ast=4$ weeks (or 28 days). To capture the impact of delayed $t^\ast$, we consider the {\bf delayed-initial} setting by which we change the initial distribution to $(S_N,S_L,S_H,I_N,I_L,I_H)=(0.93,0.4,0.02,0.01,0,0)$
wi-\\th a higher ratio of the infectious cases and both low- and high-quality information holders. The worse initial always associates with a delayed government intervention (a greater $t^\ast$), as under delayed intervention the infodemic-epidemic dynamics will be kept on their original track to evolve for a longer time, which leads to a greater infection ratio and a greater proportion of both low- and high-quality information holders.
To capture the impact of a shorter decision horizon, we consider the {\bf shorter-horizon} setting where we shrink $T^\ast$ from 4 weeks in the benchmark setting to 1 week (or 7 days) in the comparison setting. Throughout all synthetic set-ups above, the remaining model parameters are kept constant.

To align our synthetic analysis with the real-world infodemic-epidemic dynamics, we calibrate our model with the COVID-19 infection data reported in the earliest epicenter, Wuhan, China. The data collection period is from Jan. 10 2020, when the Wuhan government and the CDC in China started the daily report of COVID-19 infectious cases in Wuhan, to Jan. 24, 2020, when Wuhan was officially locked down. There was not any external intervention conducted within Wuhan during these two weeks, therefore, the epidemic spreading of COVID-19 can be approximately viewed as occurring at its natural speed within this period. Counterfactual analysis based on the model parameters trained during this period can better reflect reality. Since the entire population is susceptible to COVID-19\cite{zou2020sars,wu2020nowcasting}, $N=11$ million, the official population size in Wuhan, is taken as the base number to calculate the infection ratio. Without loss of generality, we suppose that at the initial time, all the initially reported 41 infectious cases had no information regarding COVID-19, i.e. $I_L(0)=I_H(0)=0$. Then, the initial ratio of low- and high-quality information holders, $S_L(0)$ and $S_H(0)$, together with four information parameters $\gamma$ and $\beta$, transmission rate $b_m$ and $b_h$ and the panic parameter $\theta$ are unknown and need to be calibrated with the infection data. To calibrate the unknowns, we suppose the reported infection number at day $t$ is a random number generated from the Poisson distribution with mean parameter $N\cdot I(t)$, then the unknowns can be trained as the minimum of the square-sum loss function:
\begin{align}
    loss&\left(S_L(0),S_H(0),\gamma,\beta, b_m,b_h,\theta\right)=\notag \\
    \sum_{t}&\left(\mathcal{I}(t)-I\left(t;S_L(0),S_H(0),\gamma,\beta ,b_m,b_h,\theta\right)\cdot N\right)^2
\end{align}
where $\mathcal{I}(t)$ is the reported infection number at day $t$ in the data.\\ $I\left(\cdot;S_L(0),S_H(0),\gamma,\beta,b_m,b_h,\theta\right)$ is the infection ratio trajectory that solves the ordinary differential equation system subject to the given set of parameters.

Based on the calibrated model parameters, a set of counterfactual analyses is carried out to detect whether there exists an alternative information intervention strategy that can help better contain the outbreak of COVID-19 in China. In the counterfactual world, we still focus on the two classes of strategies, information injection, and information blocking. Similar to the synthetic setting, we consider the joint adjustment of alternative injection time $\tau$ within the range $(t^\ast,T^\ast)$ and alternative blocking degree $s$ ranging from $0$ to $2$ such that
the information parameters adjusted by $s$ are given by $\gamma^s_1=\hat{\gamma}_1\cdot s,\,\gamma^s_2=\hat{\gamma}_2\cdot s$ and $\beta_1^s=sign(\hat{\beta}_1)|\hat{\beta}_1|\cdot s$ where $\hat{x}$ denotes the estimated value of parameter $x$. For the adjustment of blocking, to avoid the scale effect, we re-scale the blocking degree $s$ by the absolute value of the estimated information parameters. Due to the potential negativity of $\beta_1$, we let $sign(a)$ represent the sign of $a$ and use the multiplier $sign(\hat{\beta}_1)|\hat{\beta}_1|$ to preserve both scale and the sign of $\hat{\beta}_1$. Also analogous to the synthetic setting, we consider the impact of different decision time parameters, $(t^\ast,T^\ast)=(0,30),\,(0,180),\,(30,210)$, where the unit of time is day and day $0$ is naturally identified with the initial reporting time, Jan. 10, 2020.

For both synthetic analysis and the counterfactual analysis based on model calibration, the functional form of $g_\tau=g(H_\tau,H_\tau+L_\tau)$ is fixed as the following
\begin{equation}\label{gtau}
g_\tau=\begin{cases}
\Scale[0.85]{\left(|\gamma_1|,|\gamma_2|,|\beta_1|,|\beta_2|\right)\cdot H_\tau(L_\tau+H_\tau)}, & L_\tau+H_\tau>0.1\\
0, &else
\end{cases}
\end{equation}
where $\cdot$ stands for the product between constant and vector. The discontinuity in \eqref{gtau} captures the threshold effect that without a sufficient amount of information permeable in the system, the government even cannot notice the existence of viral infection, nor is possible to offer higher-quality information.

\section{Results}
\subsection{Government decision complexity and intervention dilemma}

\begin{figure}
    \centering
\includegraphics[width=0.9\textwidth]{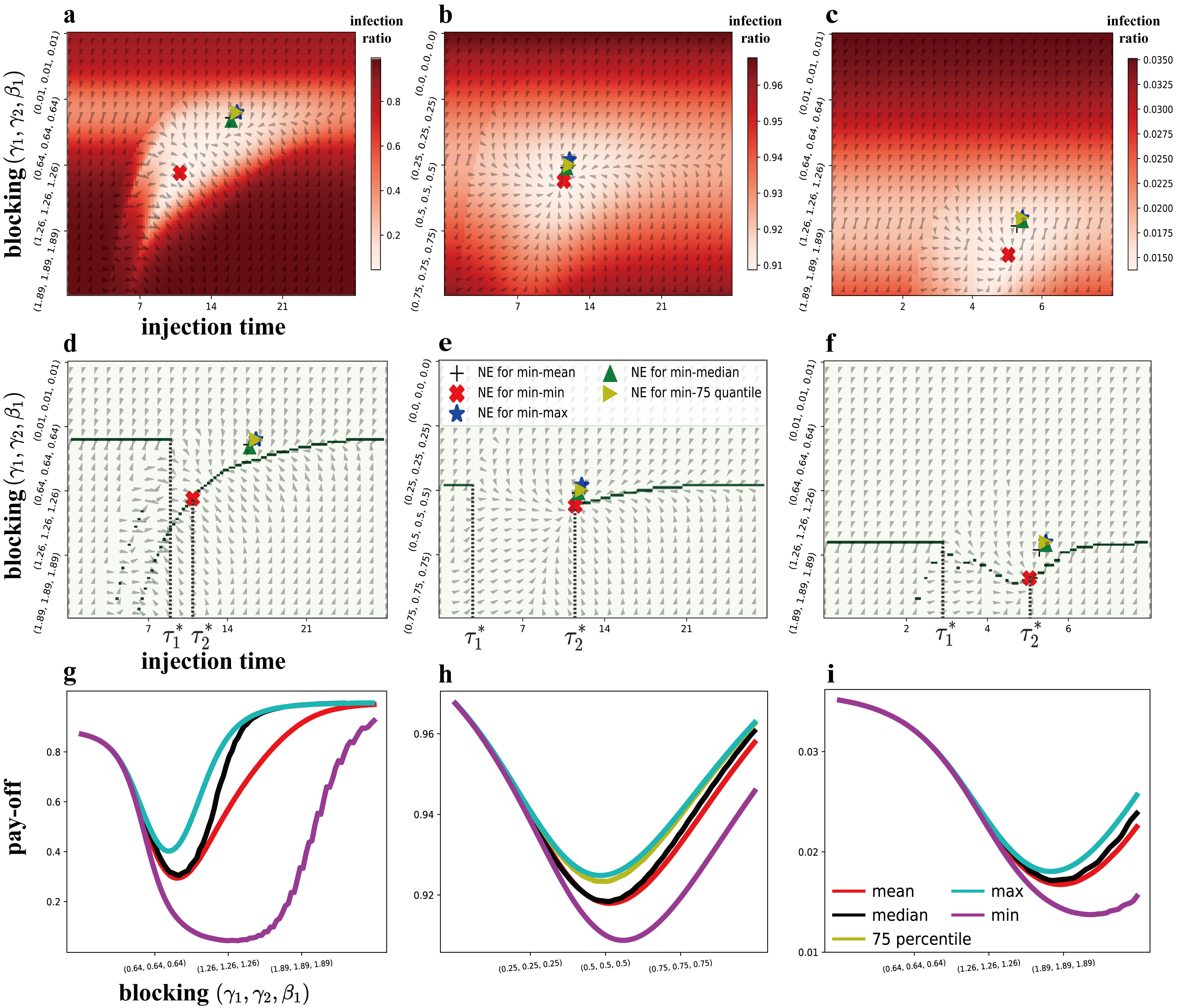}
   \caption{Joint effects of government intervention.} \floatfoot{\footnotesize {\bf (a-f)} Variation trend of final infection ratio $I(T^\ast)$ in response to the joint adjustment of blocking $s$ and injection time $\tau$. The variation trend is characterized through the gradient field of $I(T^\ast)$ with respect to $s$ and $\tau$ under fixed initial conditions. In {\bf a-f} the gradient field $\vec{\nabla}=-\left(\frac{\partial I(T^\ast)}{\partial \tau},\frac{\partial I(T^\ast)}{\partial s}\right)$ is represented as the background arrows. Plots {\bf a,d}, {\bf b,e} and {\bf c,f} associate with the simulation result for the {\bf benchmark}, {\bf delayed-initial}, {\bf shorter-horizon} setting, respectively. Throughout {\bf a-c}, we give the heat maps of $I(T^\ast)$, marking the value of $I(T^\ast)$ at each pair $(\tau,s)$ through the darkness of background color. Throughout {\bf d-f}, we mark, by dark dots(curves), those $(\tau,s)$ satisfying $\frac{\partial I(T^\ast)}{\partial s}=0$ and $\frac{\partial I(T^\ast)}{\partial \tau}\geq0$. They are the steady states under the forward-looking gradient field $\overrightarrow{\nabla}_f=-\left(\frac{\partial I(T^\ast)}{\partial \tau}\cdot \mathbf{1}\left(\frac{\partial I(T^\ast)}{\partial \tau}\geq 0\right),\frac{\partial I(T^\ast)}{\partial s}\right)$ by which the injection time can only be adjusted forwardly, the backward adjustment is not allowed. Restricting the adjustment direction of injection time reflects the fact that the decision on injection time is irreversible in time. {\bf (g-i)} Aggregated infection ratios over different injection time $\tau$ against fixed blocking degree $s$. The aggregated $I(T^\ast)$ is considered as the pay-off for the government adopting action $s$ at the initial decision time $t^\ast$ in the dynamical game discussed in the main text. The variation of this pay-off against $s$ is plotted in {\bf g-i} with respect to various aggregation functions, including the \enquote{max}, \enquote{min}, \enquote{mean}, \enquote{median} and \enquote{75-quantile} functions. {\bf g}(/{\bf h}/{\bf i}) shares the same set of initial conditions and initial decision time $t^\ast$ with {\bf a,d}(/{\bf b,e}/{\bf c,f}). Finally, in alignment with each aggregated pay-off function in {\bf g}/{\bf h}/{\bf i}, the induced Nash-equilibrium (NE) combinations of blocking degree and injection time is plotted {\bf a,d}/{\bf b,e}/{\bf c,f}.}\label{GF-intervention}
\end{figure}

Fig. \ref{GF-intervention} presents the results of our synthetic analysis. In Fig. \ref{GF-intervention} the complexity behind the government's decision on the optimal intervention strategy emerges, from which we witness quite a few interesting but counter-intuitive phenomena as below.

\vspace{0.1cm}
\noindent\textit{\textbf{Waiting period exists between the time when government becomes capable of information injection and the optimal injection time.}} The set of all steady states under the forward-looking gradient field of $I(T^\ast)$ (dark dots in Fig. \ref{GF-intervention}d-\ref{GF-intervention}f) are naturally clustered around two separated curves, where the horizontally straight line on the left is locally attractive via the gradient field (gray arrows in Fig. \ref{GF-intervention}a-\ref{GF-intervention}f). By the discontinuous design of $g_\tau$ in Sec. Method, the injection time $\tau_1^\ast$ associated with the right end of the horizontal dark line is the earliest time when information injection can effectively impact the disease dynamics. Apparently from Fig. \ref{GF-intervention}d-\ref{GF-intervention}f, a gap exists between $\tau_1^\ast$ and the optimal time $\tau_2^\ast$ associated with the red \enquote{X} in Fig. \ref{GF-intervention}a-\ref{GF-intervention}c (for easiness of comparison, the red \enquote{X} is also marked at the same location in Fig. \ref{GF-intervention}d-\ref{GF-intervention}f), where the red \enquote{X} represents the globally optimal combination of blocking degree and information injection time in terms of minimizing the final infection ratio. Also remarkably in Fig. \ref{GF-intervention}d, the gradient field, projected to the dimension of injection time, points away from $\tau_1^\ast$ to the right at $\tau_1^\ast$ and points toward $\tau_2^\ast$ from the left at $\tau_2^\ast$. This observation implies government should wait and hold the information for a gap period $\tau^\ast_2-\tau^\ast_1$, rather than release them to the public immediately
since releasing becomes feasible. The existence of a waiting period contradicts to the common belief that the best strategy for the government is to publish high-quality information as soon as possible. Although counter-intuitive, the existence of a waiting period is deeply rooted in the trade-off expressed in Fig. \ref{intervention-tradeoff}c and \ref{intervention-tradeoff}d. No waiting period suggests an earlier information injection which leads to an immediate drop-down of the current infection ratio $I(\tau+\Delta\tau)$ during the short time after injection (the gap between the green and red line in the interval from $\tau$ to $\tau'$ in Fig. \ref{intervention-tradeoff}c). A lower infection ratio at the later injection time $\tau'$ would also shrink the risk of a further outbreak, measured by the gap between the green line and red line in the horizon time $T^\ast$ in Fig. \ref{intervention-tradeoff}c. Therefore, the delayed injection induces a negative $A_1$ and a higher final infection ratio. From this perspective, we can confirm the positive immediate effect of early injection on containing virus outbreaks. However, on the other hand, compared to injection after waiting, earlier information injection cannot generate a substantial structural impact on the information propagation process, hence has limited restriction on the virus outbreak speed in the later time, leading to a positive $A_2$ in Fig. \ref{intervention-tradeoff}c and a greater final infection ratio. Such a negative structural effect offsets the positive immediate effect of early injection and leads to an overall negative impact on the final infection. The existing literature often neglects the negative structural impact induced by earlier information injection, leading to an over-optimistic attitude toward the functionality of quick information release.

\vspace{0.1cm}
\noindent\textit{\textbf{Government faces the dilemma between the optimal strategy that minimizes the cost of its own mistakes and the socially optimal strategy that minimizes the final infection ratio.}} Fig. \ref{GF-intervention}a also indicates a temporal inconsistency in the joint decision of blocking degree and injection time. In fact, the global optimality (the red \enquote{X}) in Fig. \ref{GF-intervention}a can be reached if and only if the following latent assumption holds: i) the decision of blocking degree is made simultaneously with that of information injection; ii) the government have complete information on the form of injection function $g_\tau$.
However, none of the two conditions can really hold in practice. The decision of blocking degree is always made ahead of the injection time. Meanwhile, at the decision time of blocking degree ($t^\ast$), the information in Fig. \ref{GF-intervention}a regarding how the degree would affect the effectiveness of information injection at different future injection times is completely absent for the government. The incomplete information induces a conflict. On one hand, the optimal injection time can reach its optimal containment effect if and only if the blocking can hold on a fairly low level. On the other hand, given such a low-level blocking, the infection ratio could be much worse than that on a higher blocking level in case the real injection time deviates from its theoretical optimum. This uncertainty leads to a choice dilemma
that can be perfectly reformulated as a two-player dynamical game with incomplete information, in which the first player is the government at the decision time of blocking, $t^\ast$, while the second player is the future government facing the decision of injection time. The pay-off matrix is given by Fig. \ref{GF-intervention}a, the vertical axis (blocking degree) corresponds to all feasible actions of player 1, and the horizontal axis (injection time) corresponds to all feasible actions of player 2. Player 2 has complete information on the blocking degree taken by player 1 at its decision time, therefore its pay-off is exactly equal to the infection ratio colored in Fig. \ref{GF-intervention}a. While due to incomplete information regarding the actual injection time of player 2 at $t^\ast$, for every blocking degree $s$, the pay-off for player 1 is only an aggregation of infection ratios for all different $\tau$s and the fixed $s$. If both the two government players behave rationally, the final decision on blocking degree $s$ and injection time $\tau$ should achieve a Nash equilibrium (NE). In Fig. \ref{GF-intervention}g-\ref{GF-intervention}i, we plot the aggregated pay-off for government at $t^\ast$ against the blocking degree $s$ under a variety of aggregation
principles. The resulting Nash-equilibrium joint decision on $\tau$ and $s$ are plotted in Fig. \ref{GF-intervention}a-\ref{GF-intervention}d. This game-theoretical interpretation of the government decision dilemma provides a different way to think of the failure of government in the early-stage fighting with infectious diseases. Fig. \ref{GF-intervention}d shows if government at $t^\ast$ is prudent and takes the max aggregation, i.e. following the well-known \enquote{mini-max} decision principle\cite{cabulea2004making,winterich2015disgusted}, the optimal blocking degree exactly coincides with the degree under the local steady states of the gradient field for small injection time (the horizontal dark line in Fig. \ref{GF-intervention}d), which is distant from the global optimal blocking degree associated with the red \enquote{X} in Fig. \ref{GF-intervention}a and \ref{GF-intervention}d. In real-world catastrophes, such as the pandemic of SARS, H1N1, and COVID-19, the \enquote{mini-max} principle is more likely to be adopted for government decisions. Given a blocking degree $s$, the maximally aggregated infection ratio provides an upper bound for the government's mistakes in selecting injection time. The \enquote{mini-max} principle attempts to minimize this upper bound, which is equivalent to minimizing the responsibility of government. In contrast, the global optimal intervention strategy is equivalent to the Nash-equilibrium under the \enquote{mini-min} principle, which might be optimal for the entire society as it minimizes the final infection ratio. But once the government makes mistakes in the second-stage decision for injection time, the potential loss will be giant. For instance, in Fig. \ref{GF-intervention}g, at the blocking degree $s^\ast$ associated with the minimum of the \enquote{min} curve, the difference in the infection ratio between government correctly and mistakenly selecting the injection time can exceed 70\% in the worst case (measured by the difference between the \enquote{max} and \enquote{min} curves at $s^\ast$ in Fig. \ref{GF-intervention}g). Such a huge difference in infection ratio means a giant cost for the government's mistake, which is not affordable for the government in the initial decision timing $t^\ast$. Consequently, the optimal intervention strategy
in the views of government has the natural tendency to deviate from the global optimal strategy, which partially explains the universal loss of control of the COVID-19 pandemic by the central government in most major countries around the world, such as China, US, Italy, Great Britain and the like.

\vspace{0.1cm}
\noindent\textit{\textbf{Delayed initial decision time $t^\ast$ relieves government dilemma but leads to a much higher infection ratio.}} Compared to Fig. \ref{GF-intervention}a, \ref{GF-intervention}d and \ref{GF-intervention}g, government decision under a delayed initial time $t^\ast$ is plotted in Fig. \ref{GF-intervention}b, \ref{GF-intervention}e and \ref{GF-intervention}h. Due to the delay, there are more infectious cases and more low-quality information holders in society at the beginning. In the real world, it always takes time to figure out the occurrence of the crisis, hence government can never be in a decision position until the outbreak has lasted for a while. Fig. \ref{GF-intervention}b, \ref{GF-intervention}e and \ref{GF-intervention}h capture the challenges in the delayed decision setting.
Comparing Fig. \ref{GF-intervention}b and \ref{GF-intervention}e with Fig. \ref{GF-intervention}a and \ref{GF-intervention}d, several structural changes of the gradient fields can be observed. First, the waiting period is even prolonged under the delayed $t^\ast$, reflecting that given more infections and more low-quality information holders, the structural impact of information injection captured by $A_2$ in Fig. \ref{intervention-tradeoff}c becomes even more prevalent compared to the effect of $A_1$ on the immediate reduction of infection number. On the other hand, the temporal inconsistency between the decision of blocking degree and injection time gets weakened in the delayed case which reflects as the difference shrinks between the pay-off for government at $t^\ast$ under different aggregation methods (in Fig. \ref{GF-intervention}h), as well as the fact that the red \enquote{X} in Fig. \ref{GF-intervention}b and \ref{GF-intervention}e is getting closer to the Nash-equilibrium strategies under the other pay-off aggregation methods, especially under \enquote{mini-max} principle. The decreasing temporal inconsistency reduces the uncertainty and the cost of making mistakes faced by government and makes the decision on blocking degree much easier. But the subsequent side effect is significant as well, the minimal infection ratio in Fig. \ref{GF-intervention}b is much higher than that in
Fig. \ref{GF-intervention}a, reflecting by the different scale of the color bar associated with Fig. \ref{GF-intervention}a and \ref{GF-intervention}b, respectively. In Fig. \ref{GF-intervention}a, the global minimal infection ratio can be maintained below 20\%, while in Fig. \ref{GF-intervention}b, the global minimal infection ratio exceeds 90\% which is 70\% higher than that in the former case.
As the worse final infection ratio results from the corresponding initial condition induced by the delayed $t^\ast$, the comparison between the first and second column of Fig. \ref{GF-intervention} reveals a social dilemma faced by government. On one hand, a delayed $t^\ast$ can help reduce the inconsistency between the optimal strategy for government measured by minimizing the cost of making mistakes and the optimal strategy for the entire society measured by minimizing the final infection ratio, hence government may prefer to delay so as to reduce the difficulty of making the correct decision. However, on the other hand, a delayed $t^\ast$ would make the \enquote{consistently optimal} strategy no longer \enquote{good} via significantly rising up the final infection ratio and causing more social welfare loss.

\vspace{0.1cm}
\noindent\textit{\textbf{Shortening decision horizon $T^\ast$ relieves government dilemma at the cost of a much higher infection ratio.}} Similarly to delaying $t^\ast$, Fig. \ref{GF-intervention}c, \ref{GF-intervention}f and \ref{GF-intervention}i show that the variation on the decision horizon $T^\ast$ can also significantly impact the trade-off between the difficulty of government decision and social optimal infection ratio. Compared to Fig. \ref{GF-intervention}a, the decision horizon $T^\ast$ in Fig. \ref{GF-intervention}c is compressed from 4 weeks to only 1 week. The severity of the dilemma is also reduced in terms of the difference in optimal blocking degree under different aggregation principles. In Fig. \ref{GF-intervention}f, the difference in the optimal blocking degree under the \enquote{mini-min} aggregation principle and under the \enquote{mini-max} aggregation principle is 0.35, which is significantly lower than the difference (0.54) in Fig. \ref{GF-intervention}d. Meanwhile, the optimal blocking degrees in Fig. \ref{GF-intervention}f under different aggregations are universally lower than those in Fig. \ref{GF-intervention}d, reflecting that the horizontal dark line in Fig. \ref{GF-intervention}f is uniformly below that in Fig. \ref{GF-intervention}d. In the other words, viewed in a shorter horizon, a lower final infection ratio can be achieved via a much freer information environment, which agrees with the widely held belief on that free information propagation suppresses disease spreading\cite{kim2019incorporating}. But if we compare the relative position of two red \enquote{X}s and the associated infection ratio in Fig. \ref{GF-intervention}a, a novel temporal inconsistency appears. The global optimal strategy viewed in a shorter decision horizon $T^\ast$ leads to an extraordinarily
high infection ratio ($>80\%$) when the horizon is extended. In fact, a freer information environment helps sharply rise up the proportion of high-quality information holders in the short run, which reduces the infection ratio within this group. But in a longer horizon, if the trend of information decay is not reversed, \enquote{holding high-quality information} is only a temporal status. As time passes away, the initially high-quality information holders will ultimately decay into low-quality information holders, which significantly enlarges their infection risk. This observation implies that the decision horizon is critical to the effectiveness of optimal intervention strategies. If the government is myopic and puts overweight on containing the disease outbreak in the short run, it is likely to be misled by the myopic preference and make the wrong decision that causes huge social welfare loss viewed in a relatively longer time horizon.

\subsection{Case study of COVID-19 pandemic in mainland China}
\begin{figure}
    \centering
\includegraphics[width=0.9\textwidth]{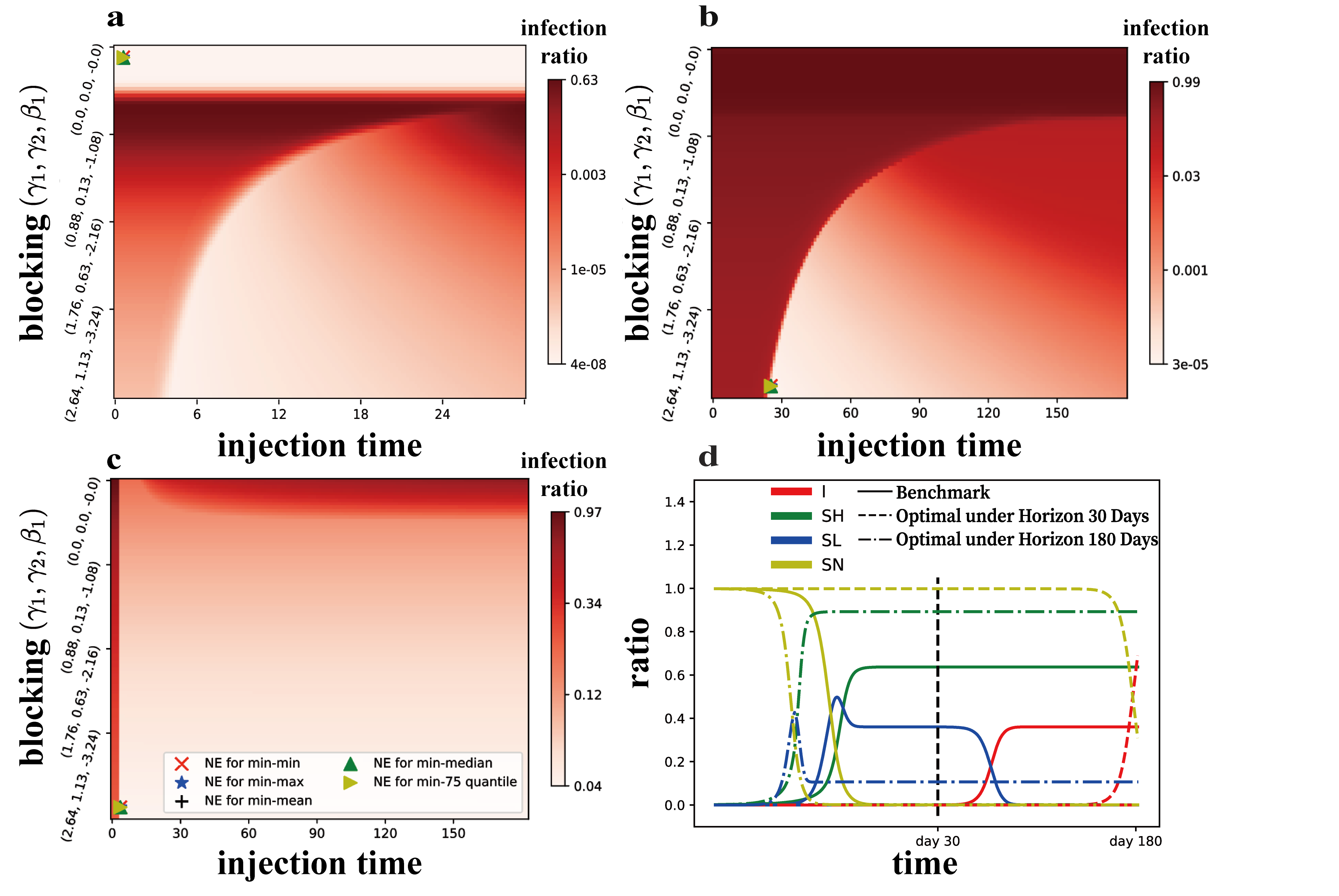}
     \caption{Response dilemma during real-world pandemics.}
     \floatfoot{\footnotesize {\bf (a-c)} Heat maps of the final infection  ratio $I(T^\ast)$. The heat map sketches the joint trade-offs faced by the Chinese government when dealing with information control during the COVID-19 pandemic. {\bf (a)} Benchmark setting where the blocking decision timing $t^\ast$ is set as the initial time of the data, i.e. Jan. 10, 2020, the decision horizon $T^\ast$ is 30 (days), i.e. one month. {\bf (b)} Decision under a longer horizon with $(t^\ast,T^\ast)=(0,180)$. {\bf (c)} Decision under delayed initial decision time and longer horizon with $(t^\ast,T^\ast)=(15,195)$. Plot {\bf b,c} reveal the impact of the variation of $t^\ast$ and $T^\ast$ on government choice dilemma. Plot {\bf d} sketches the evolutionary trajectories of the fractions $I$, $S_N$, $S_L$, and $S_H$ during the first 180 days under calibrated model parameters and under the globally optimal intervention strategies in {\bf a} and {\bf b} respectively. All the counterfactual results in {\bf a-d} are calculated on the basis of the calibrated model parameters.  The estimated information parameters $\hat{\gamma_1}=1.76$, $\hat{\gamma}_2=0.63$, $\hat{\beta}_1=-2.16$ and $\hat{\beta}_2=1.23$, the transmission rate $\hat{b}_m=0.02$, $\hat{b}_h=2.94$, the panic parameter $\hat{\theta}=0.97$ and the initial $\hat{S}_L(0)=5\times 10^{-6}$, $\hat{S}_H(0)=1.3\times 10^{-3}$. Based on calibration results and the counterfactual set-ups, the government intervenes in the infodemic-epidemic dynamics via information blocking and injection, and the execution of intervention strategies is identical to that in the synthetic case discussed in Fig. \ref{GF-intervention}.  }\label{dilemma china}
\end{figure}

The model calibration result and the counterfactual analysis based on calibration are reported in Fig. \ref{dilemma china}, which provides us with some new insights into the dilem-ma that are not covered in the synthetic settings but the government has to face in the real world.

\noindent\textit{\textbf{Completely blocking information spreading is the universally optimal strategy for the Wuhan government under myopic decision horizon.}} Surprisingly from Fig. \ref{dilemma china}a, the Chinese government didn't face to the dilemma if there is no delay and the decision horizon $T^\ast$ is set to 30 days. The optimal blocking should be implemented to its strongest degree, i.e. cut off all contacts that spread information regarding COVID-19 no matter whether or not the information content is high-quality. Meanwhile, the optimal injection time is zero, i.e. no information injection at all. Remarkably, such a combination of blocking degree and injection time is universally optimal for all different ways of aggregating the intervention pay-off. Especially, the \enquote{mini-min} and \enquote{min-max} principles agree with each other. This surprising result can be explained through the difference between the transmission rate under different information statuses ( $\hat{b}_h/\hat{b}_m\approx 300$), and the high conversion rate to low-quality information from no information state ($\hat{\gamma}_1=1.76$), as well as from information decay ($\hat{\beta}_2=1.22$). In fact, the high ratio $\hat{b}_h/\hat{b}_m$ implies panic-induced infection is the dominant driving force for the accumulation of infectious cases. Since panic happens to and only to low-quality information holders, then a high-conversion rate from the other information status to low-quality status makes the whole population suffer from a high risk of panic-induced infection. Comparatively, the infection risk due to no awareness is relatively low. Consequently, the optimal strategy for the government is to keep everyone unaware of COVID-19, in exchange for a low transmission rate $\hat{b}_m$.

\vspace{0.1cm}
\noindent\textit{\textbf{Government dilemma appears as the polarized optimal intervention strategies under different decision horizons.}} The \enquote{keeping unawareness} strategy is only temporally optimal. If the decision horizon $T^\ast$ is extended 180 days (Fig. \ref{dilemma china}b) and/or the initial decision timing $t^\ast$ is delayed (Fig. \ref{dilemma china}c), a restriction-free information environment should be maintained under the optimal, and the optimality is concurred by all different aggregation principles in Fig. \ref{dilemma china}b and \ref{dilemma china}c, especially by both the \enquote{mini-max} and \enquote{mini-min} principles. Also remarkably, if the optimal blocking under a 30-day horizon is actually executed, the final infection ratio viewed under the 180-day horizon is terribly high ($>$60\%) which is more than 60\% higher than the infection ratio reached by the optimal strategy under the 180-day horizon, no matter whether the initial decision is delayed. These observations suggest that in the real world, the choice dilemma that the Chinese government faces no longer comes from the inconsistent temporal decision induced by different aggregation principles and different risk preference behind government decision, but arise from the temporally inconsistent effect of the same set of strategy viewed from different time horizons. The temporally inconsistent containment effect is deeply rooted in the complexity of the coupled infodemic-epidemic dynamics. Comparing the half-a-year horizon with the one-month horizon, the difference in the optimal blocking degree comes mainly from that in the one-month case, due to the low initial amount of infectious cases, the disease dynamics are mainly driven by the high transmission rate $\hat{b}_h$ for low-quality information holders. But as time pass away, as shown in Fig. \ref{dilemma china}d, the infectious cases keep accumulating meanwhile the number of healthy low-information holders are decreasing (reflected as the decrease of the blue lines in a later time in Fig. \ref{dilemma china}d), in the second stage of disease dynamics, the infection will be mainly driven by the infection of healthy but unaware individuals (reflecting as the sharp decrease of the dash yellow line accompanied with the sharp increase of the dash red line in Fig. \ref{dilemma china}d). The switch of driving force increases the effectiveness of free information propagation and information injection while suppressing the effect of blocking, which makes government once again face a dilemma between myopic decisions versus long-horizon decisions. As documented in the literature of behavioral economics and psychology, human beings are naturally inclined to over-weigh the current loss and pretend to be blind to the loss that happened in the longer future\cite{benartzi1995myopic,tversky1974judgment,kahneman2013prospect}. On the other hand, making decisions for a longer horizon is a much more complicated task that needs more information to overcome the increased uncertainty. At the initial decision time, information is too limited to support such a longer-horizon decision. Therefore, driven by both the myopic preference and initial information limitation, government when facing the contradicted optimal strategy in Fig. \ref{dilemma china}a and Fig. \ref{dilemma china}c is more likely to select the short-run optimal strategy in Fig. \ref{dilemma china}a, rather than the long-run optimal in Fig. \ref{dilemma china}c, which makes the society suffer from a much higher infection ratio. In the real world, the initial reaction of the government in the epicenter, Wuhan, followed exactly the myopic optimal strategy, it did not disclose any information to the public, meanwhile hid all key information regarding the transmissibility of COVID-19 and prohibited the public discussion on the internet. As known, this short-run optimal reaction leads to the outbreak of COVID-19 in China. The evidence of Wuhan demonstrates that the myopic decision mode and the induced inaccurate evaluation on the long-term infection ratio form a driving force to mislead the government's choice of the correct intervention strategy and cause severe social welfare loss. The same logic applies as well to understand the failure of many other countries in selecting the correct intervention strategy during the early-stage containment of COVID-19, such as the US.

\subsection{Fighting with government dilemma from the view of network}
\begin{figure}
\centering
\includegraphics[width=0.9\textwidth]{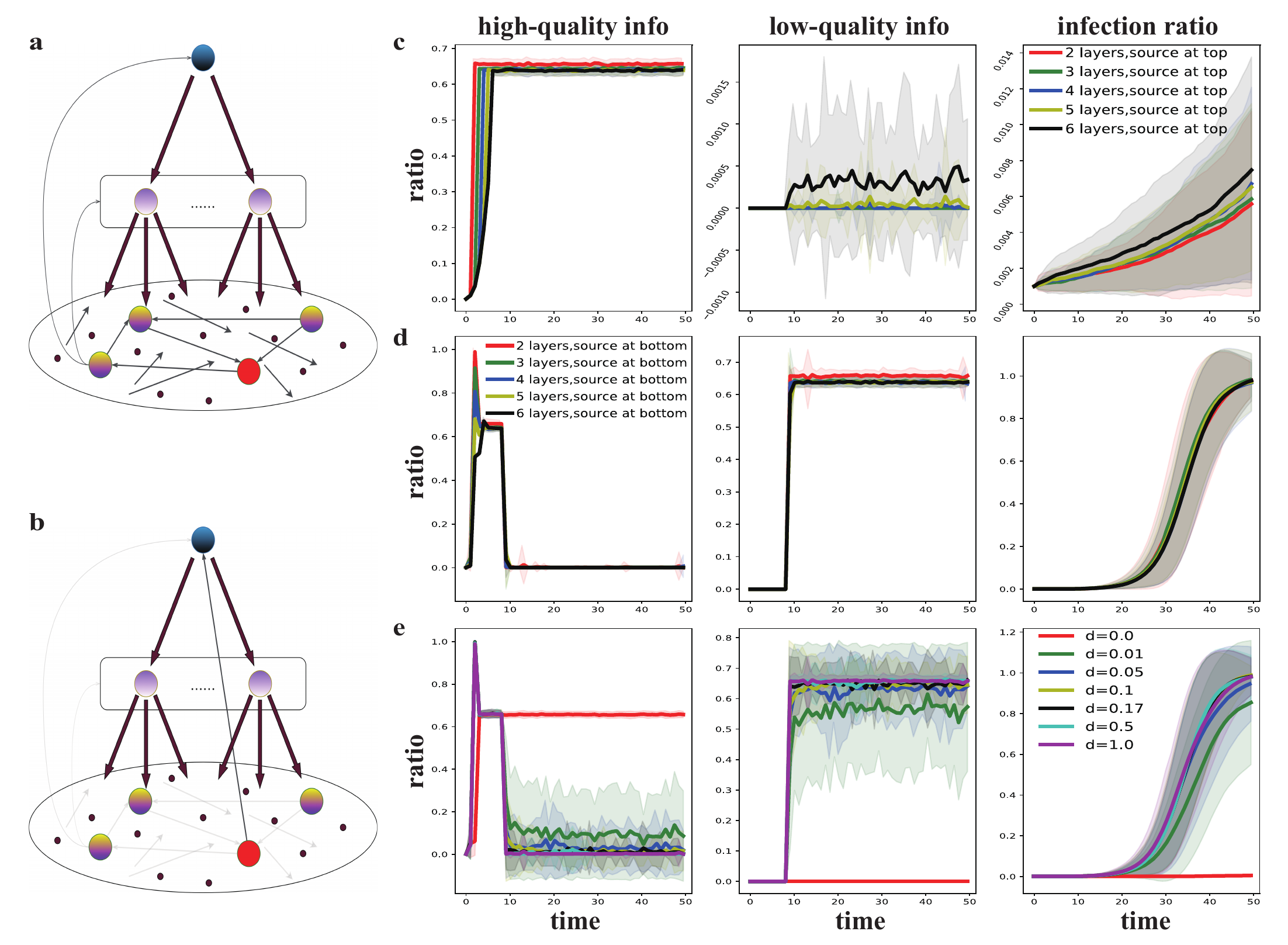}
     \caption{Information dynamics on a hierarchical network and the effect of deloop.}
     \floatfoot{\footnotesize {\bf (a)} Example of the hierarchical network structure (with three layers). The number of nodes in each layer increases along with a rank decrease of the layer within the hierarchy. Nodes in the top 2 layers are extraordinarily influential to those in the next-level layers (reflecting as the thick arrows pointed downward from the high to the lower layers in {\bf a}), but nodes in the bottom layer only have weak influence on each other and a much weaker influence to the nodes in the top 2 layers (reflecting as the thin arrows within the third layer and pointed upward from the third layer in {\bf a}). Such a hierarchical network structure captures the information propagation mechanism among the central government (top layer), local government (intermediate layer), and the public (bottom layer). At the bottom layer, the red dot represents the source information holder at the initial time. Due to the lack of a direct link from the source to the central government, it takes much more time and more quality loss before the information can be delivered to the central government and passed to the public. The time and quality loss facilitates the rumor spreading via the low-layer links and loops within the bottom layer. {\bf (b)} Illustration of how direct warning system and deloop can reshape the network structure and affect the decay process. The direct warning system works as adding an arrow pointing toward the top from the information source, while the deloop works as cutting off all links that facilitate loop formation within the bottom layers. Deloop together with a direct warning system makes the resulting network a four-layers hierarchical network, where the top layer is replaced with the source information holder. The renewed network structure stops the decay as shown in {\bf e}. {\bf (c-d)} Trend of quality decay and infection growth within a hierarchical information network with a different number of layers and different initial assignments of source information. The number of nodes in {\bf c} and {\bf d} is fixed to 1000, the number of layers varies from 3 to 7, and the link structure of the network is given by a random weighted adjacency matrix with the out-degree weight distribution determined by the real out-degree distribution of the follower-ship network on Weibo, the biggest Chinese online social media. In {\bf c} the source information is always assigned to the unique top-layer node, while in {\bf d} assigned randomly to one node in the bottom layer. {\bf (e)} Effect of deloop on mitigating the trend of quality decay and infection. To measure the degree of deloop, we set a parameter $d\in[0,1]$. In {\bf e}, deloop works as changing the entry weight of the adjacency matrix $W$ by multiplying $d$ with $W_{ij}$ for those $ij$s that form loops in the network. $d=1$ means no deloop, $d=0$ implies complete deloop described in {\bf b}. For robustness, the simulation result in {\bf c-e} is averaged over 100 realizations for each fixed setting.
     }\label{deloop}
\end{figure}

In the aforementioned analysis, the dilemma faced by the government relies on information quality decay, i.e. $\beta_1> 0$ and/or $\beta_2>0$. To verify this premise,
in this section, we present an example network structure through which fast information decay is unavoidable. Meanwhile, we discuss the potential re-structure of the network that helps relieve the government from the dilemma. We let the information dynamics occur through a hierarchical network, the network topology is illustrated in Fig. \ref{deloop}a. The key assumption regarding the information dynamics is that every node, when receiving information from the others, updates his/her own quality level via the weighted sum of the quality of all received information with the weight of each delivery node determined by their out-degree (a rigorous mathematical formulation of propagation mechanism is provided in the Appendix). The out-degree of every node provides a measure of its influence on the information propagation process, which is a widely used assumption in the literature\cite{liu2016characterizing,cha2009measurement} of propagation dynamics, such as the virtual dynamics behind the calculation of the renowned Page Rank \cite{page1999pagerank}.  The hierarchical network topology is quite realistic in describing information propagation during the pandemic because, unlike the propagation of general news, the government always plays a central role in the propagation of disease-related information. The extraordinary influence of government on the propagation process naturally leads to an inequality between itself and the public, resulting in the hierarchical structure of the network. We show in Fig. \ref{deloop}c and \ref{deloop}d that in a hierarchical network, the overall quality level of the system is hyper-sensitive to both the number of layers in the hierarchy and the distribution of the source information holders. If the source information holders are on the top level, i.e. the government itself holds the source information, the quality level of the whole system can be maintained at a relatively high level which corresponds to a low infection ratio in the end. But the increasing number of layers lying between the government and the bottom-layer public would offset the advantage of the government holding source information as shown in Fig. \ref{deloop}c, reflecting the importance of a flattened management structure during a disease pandemic. On the contrary, if the source information is held by some nodes at the bottom layers (i.e. the individuals), the quality level would decay very fast, as shown in Fig. \ref{deloop}d. This result suggests that in a hierarchical society where information propagation channels are by and large shaped by the hierarchical social structure, the coupled dynamics of information quality decay and infectious disease outbreak would be accelerated. This finding reveals a deep connection between the structure of a society and its vulnerability to infectious diseases.

To restrict the negative effect of the hierarchical network on the infodemic-epidemic dynamics, we suggest as in Fig. \ref{deloop}b that i) strengthening the links from source information holders to the top-layer government; ii) cutting off those links that do not affect the connectivity of the entire network meanwhile can reduce the number of loops in the network ({\it deloop}). Combining i) and ii) can significantly decelerate the coupled dynamics of information decay and disease outbreak (see Fig. \ref{deloop}e). Method i) is equivalent to setting a direct warning system for risky infectious diseases, which helps flatten the hierarchical structure and reduce the quality decay during propagation from the bottom to the top. Although such a warning system has been established in most major countries around the world, as we have seen during the pandemic of COVID-19, it fails in most countries to send an effective warning signal to the central government. The failure can be partially explained by the absence of the second method, {\it deloop}. In fact, the existence of loops speeds up the propagation of low-quality information. Without a deloop, a direct warning system can only guarantee that the government can receive information with very high quality in the very beginning, it has nothing to do with the information quality decay via rumors circulating around loops. Consequently, the information decay might be delayed to some extent as shown in Fig. \ref{deloop}e, but it is ultimately unavoidable together with a very high infection ratio at the end. This suggests the importance and complexity of information management during a public health crisis, such as the pandemic of COVID-19.

The effectiveness of deloop and the connection between information circulating over loops and its quality decay also inspire us to re-think the relationship between critical public health events, such as the COVID-19 pandemic, and the booming of social media in the current era. In fact, the development of online social media, such as Twitter and Facebook, extremely extends the freedom of individuals to join the information propagation process, making it possible to freely announce personal opinions to the whole world for every single person. Social media adds tremendous loops to the underlying information network, these added loops might be good for society without concerning the pandemic of infectious disease. But in the special time of the COVID-19 pandemic, they may enlarge the risk of information decay and rumor prevailing, which ultimately facilitates the spreading of COVID-19. Therefore, the discussion in this section points out a severe conflict between the increasingly free information environment and the enlarged risk of loop-forming and the COVID-19 pandemic, which is rarely mentioned in the literature and should be put more attention in future studies.

\section{Conclusion \& Discussion}
\subsection*{Summary of major findings}

This study sophisticates the "remedy and overkill" response dilemma by modeling the effectiveness of an intervention that is restricted by information dynamics. To quickly release high-quality information to the public and maintain the overall quality level of information among the public, the government needs a fast-propagation information market to accelerate information collection and pre-processing, while that increases the risk of faster quality decay and rumor prevalence. A more restricted information market can definitely reduce the risk of rumor-induced panic, but it also reduces the space for high-quality information which will make it less likely for the government to gather and release high-quality information. The trade-off on the information propagation speed traps the government into a choice dilemma, a restriction-free information market may lead to a higher risk of panic-induced-infection in a short time, while a restricted information market compresses the chance of the government releasing high-quality information and increase the risk of long-run infection.

Facing this choice dilemma, the government has the motive to delay the intervention decision and pretend to be \enquote{blind} of the diseases pandemic, as the severity of the dilemma is declined as the infection ratio has already reached a fairly high level meanwhile the overall information quality level in the system becomes stable. The less severity of choice inconsistency means that delayed intervention will incur fewer criticisms which might be optimal for the government but deviate away from the social optimal as it induces a significant rise-up in the long-term infection ratio. Therefore, one extra social dilemma arises for the government between the politically optimal and socially optimal intervention timing.

Similar to delayed intervention, a myopic decision pattern can also reduce the severity of the choice dilemma faced by the government, because the disease dynamics might be driven by completely different forces in a relatively shorter horizon in contrast to in a longer horizon. Then, the reduced decision difficulty leads to \enquote{mistake} intervention strategy and a much greater final infection ratio and social welfare loss. We remark that compared to delayed intervention, a myopic decision might be riskier for the treatment of a public health crisis, such as the pandemic of COVID-19. It has been widely studied in psychological literature \cite{benartzi1995myopic,tversky1974judgment,kahneman2013prospect} that people are more likely to put overweight to the current and/or short-period loss than the long-period yield when facing cross-period decision under uncertainty. Therefore, the government is naturally inclined to be a myopic decision maker when facing the pandemic of viral infection and prefers the intervention strategy that is optimal for a short horizon, despite its detrimental ex-post loss viewed in the long horizon. This fact also explains the real-world failure of early-stage containment of viral infections, such as SRAR, H1N1, and the most recent COVID-19, for most major countries around the world.

 In summary, we proposed a novel model to explain the deep systematic failure rooted in the information/b-
 ehavioral heterogeneity and its induced dilemma behind government intervention decisions on COVID-19. The study highlights the complexity, information heterogeneity, and disease dynamics, and thus calls for more collaborations from social science and global health experts to help countries in designing more effective crisis management systems.

\subsection*{Limitations and future studies}

This paper has some limitations. First, we only focus on the information intervention. In the real world, there exist many other non-pharmaceutical intervention (NPI) strategies, such as social distancing and conditional quarantine \cite{zhang2020evaluating}, which are not considered in this study. To this point, we claim that most parts of the decision dilemma discussed in this study still apply to the other NPI strategies because the effectiveness of NPI strategies is also time-dependent and/or information dependent. In fact, both the NPIs and information injection attempt to convert high-susceptible individuals $S_N/S_L$ to low-susceptible $S_H$ so as to lower the population-level transmission rate into infection\cite{joel2020}. Information injection requires the government to own high-quality information before announcement which takes time and positively depends on the ratio of informed people $L+H$ and high-quality information $H$. NPI strategies are effective only if the public chooses to cooperate \cite{reluga2010game,chen2012mathematical}, the cooperativeness of the public also takes time and increases with the information amount ($L+H$) and quality level $H$. Hence, NPIs can be identified with information injection as a special type of information intervention in our framework, the choice dilemma government faces between blocking and injection holds for NPIs as well. Second, the findings of this study apply to the early stage of an epidemic crisis. As time passes away, the public will ultimately get sufficient information about the disease when the difference between the different information groups and their behavioral responses would disappear. At that stage, the infodemic may no longer be a great concern. However, on the other hand, we want to highlight that even though the trade-off of information intervention matters only for a relatively short period, its induced social-economic impact could be long-lasting, therefore its importance should not be neglected.

The current study can also be extended in the following directions. First, we focus only on the single direction that low-quality information could induce panic and higher infection risk. But reversely, the panic and fear of the disease can also impact the information propagation process and change the relative speed of propagation for different types of information. This kind of two-way feedback between information and disease dynamics induced by panic should be added into consideration. Second, combining the coupled dynamics of information and disease with an adaptive network has attracted wide attention \cite{wang2015coupled} where nodes can select their local link structure of the information network or the infection network in adaptive to the change of information or infection status, which should be a promising direction to extend the current discussion on the government response dilemma.

\begin{appendix}
\section*{Appendix A: Notation List}

\begin{table}[H]
\centering
\caption{Notations in SI-NLH model}
\label{table: table1}
\resizebox{9cm}{3cm}{%
\begin{tabular}{c|c|l}
\hline\hline
Types& Notations & Meaning\tabularnewline
\hline
\multirow{11}{*}{\makecell{Endogeneous\\ Variables}}&$S_N$&  \makecell[l]{the group of susceptible individuals without information}\\
\cline{2-3}
&$S_L$& \makecell[l]{the group of susceptible individuals with low-quality information}\\
\cline{2-3}
&$S_H$& the group of susceptible individuals with high-quality information \\
\cline{2-3}
&$I_N$& the group of infected individuals without information\\
\cline{2-3}
&$I_L$& the group of infected individuals with low-quality information \\
\cline{2-3}
&$I_H$&  the group of infected individuals with high-quality information\\
\cline{2-3}
&$I$&  the group of infected individuals\\
\cline{2-3}
&$S$& the group of susceptible individuals\\
\cline{2-3}
&$N$& the group of individuals without information\\
\cline{2-3}
&$L$& the group of individuals with low-quality information\\
\cline{2-3}
&$H$& the group of individuals with high-quality information\\
\hline
\multirow{4}{*}{Parameters}& $\gamma$&$:=(\gamma_1,\gamma_2)$ the rate of transition from group $N$ to $L$ and $H$, respectively\\
\cline{2-3}
& $\beta$&$:=(\beta_1,\beta_2)$ the rate of transition from between group $L$ and $H$\\
\cline{2-3}
& $b$&\makecell[l]{$:=(b_m,b_h)$ the rate of transmission to disease for susceptible individuals \\ under unawareness ($b_m$) and panic ($b_h$)}\\
\cline{2-3}
&$\theta$& proportion of individuals being panic within the group of susceptible\\
\hline
\hline
\end{tabular}%
}
\end{table}
\color{black}

\section*{Appendix B: Information propagation and decay in network}
The information propagation (decay) dynamics considered in this study is essentially the virtual dynamics behind the calculation of Page Rank \cite{page1999pagerank}. To describe the dynamics on a given network, we firstly define $\zeta\in[0,1]$ as the initial quality level of information for source information holders, $\delta\in(0,1)$ as the one-period decay rate of information quality, and $\eta\in [0,1]$ as the quality threshold for individuals such that the $i$th individual will be classified into $H$ at time $t$ iff at $t$ its information quality level $\zeta_{t,i}>\eta$, classified into $L$ iff $0<\zeta_{t,i}\leq \eta$, and classified into $N$ if $\zeta_{t,i}=0$. Given a network represented by its binary adjacency matrix $W$, the evolution of information group size $N$, $L$, and $H$ is governed by the following propagation dynamics on network $W$, where at time $t=0$, randomly select a set of individuals $\mathcal{I}_0^\ast\subset\{1,\dots,n\}$ as the source information holders and assign them with the information with quality $\zeta$, which yields an $n$ dimension vector $\zeta\cdot I_0$ where $I_0$ is a binary-valued vector with all entry $j\in \mathcal{I}_0^\ast$ taking the value $1$
and the other entries taking the value $0$. Then, for each time $t>0$, the information quality $\zeta_{t,i}$ for the $i$th individual is determined inductively via the following updating rule:
\begin{equation}\label{decay dynamics}
    \zeta_{t,i}=\frac{\delta}{n}\cdot\left\langle rand\left(\frac{ W_{i,\cdot}}{\langle \tilde{W}_{i,\cdot},\zeta_{t-1}^+\rangle}\right),\zeta_{t-1} \right\rangle
\end{equation}
where $\langle\cdot,\cdot\rangle$ is the inner product of two vectors, $\zeta_{t}$ is the vector formed by $(\zeta_{t,i},\dots,\zeta_{t,n})$, $\zeta_{t}^+$ is the binary vector derived from $\zeta_{t}$ in which all the entry $j$ with $\zeta_{t,j}>0$ takes value $1$, the other entries are 0, $W_{i,\cdot}$ is the $i$th row vector of matrix $W$. $rand(V)$  is a random binary vector drawn from the probability vector $V=(V_1,\dots,V_n)$ ($V_i\geq 0$ and $\sum_iV_i=1$) where the $i$th entry take value $1$ in the probability $V_i$. $\tilde{W}$ is the matrix derived from $W$ via $\tilde{W}=diag(W\cdot \mathbf{1}_n)^{-1} W$ where $\mathbf{1}_n$ is the $n$-dimensional vector with all entries taking value $1$, $diag(\cdot)$ is the operation that converts an $n$-dimensional vector to a diagonal matrix with the diagonal entries identified with the input vector. The constructions of $\tilde{W}$ and $\zeta_{t}^+$ imply that under the information propagation dynamics \eqref{decay dynamics} the information quality of individual $i$ at time $t$ is the natural decay rate $\delta$ times a weighted average of information quality of its network neighbors according to the relative network influence of each neighbor measured by their out-degree. The influence-based information propagation mechanism is widely used in literature \cite{liu2016characterizing,cha2009measurement}. From $\zeta^{t}$, the fraction of $N$, $L$ and $H$ can be calculated easily via comparing each $\zeta_{t,i}$ and the threshold $\eta$ as discussed before.

Given the information dynamics on a subtly structured network, the disease dynamics are still simulated based on the assumption of a well-mixed population. In the other words, the conversion from $S_N$, $S_L$, and $S_H$ to $I$ still follows equation \eqref{toy math}. In the agent-based simulation, this conversion is equivalent to updating a binary vector $I_t$ that stores the infection status of every node for every time $t$, and can be realized via the following steps given the initial $I_0$:
\begin{itemize}
    \item[{\bf Step 1}:] Given node $i$, if $i$ has been infected, directly continue to node $i+1$; otherwise, randomly match $i$ with another node $j\in\{1,\dots,n\}$ and $j\not=i$.
    \item[{\bf Step 2}:] If $j$ has not been infected, directly continue to node $i+1$; otherwise, determine the information status of $i$ according to $\zeta_{t,i}$ and $\eta$.
    \item[{\bf Step 3}:] if $i$ is in $S_H$, directly continue to node $i+1$; if $i$ is in $S_N$, update $i$'s infection status to $I_{t,i}=1$ by the probability $b_m$; if $i$ is in $S_L$, randomly assign $b_h$ to $i$ in the probability of $\frac{\sum{l=1}^{n}I_{t-1,l}+\theta\sum_{l=1}^{n}\mathbf{1}(\zeta_{t,l}<\eta)}{n}$; if $b_h$ is assigned to $i$, update $i$'s infection status to $I_{t,i}=1$ by $b_h$, otherwise, update $i$'s infection status to $I_{t,i}=1$ by $b_m$.
\end{itemize}
The three-step algorithm will be used to generate the infection ratio in Fig. \ref{deloop}c - \ref{deloop}e.

\end{appendix}

%
%

\section*{acknowledgements}
This work was partially supported by the National Natural Science Foundation of China (Grant Nos. 72101268 and 61673151), the Natural Science Foundation of Zhejiang Province (Grant Nos. LR18A050001),  the Major Project of The National Social Science Fund of China (Grant No. 19ZDA324), the China Postdoctoral Science Foundation (Grant No. 2021M703569), and the Fundamental Research Funds for the Central Universities (ZJU and Grant No. 224220S30024 at SEU).

%
\section*{Conflict of interest}
We declare no competing interests.

\bibliography{1.bib}

%
%

\end{document}